# Tunneling Spectroscopy of Disordered Two-Dimensional Electron Gas in the Quantum Hall Regime

Gilad Barak[1], Amir Yacoby[1], Yigal Meir[2]
[1]Department of Physics, Harvard University, Cambridge MA, 02138, USA.
[2] Department of Physics, Ben Gurion University, Beer Sheva 84105, Israel.

**Recently, Dial *et al.* presented measurements of the tunneling density of states into the bulk of a two dimensional electron gas under strong magnetic fields. Several high energy features appear in the measured spectrum showing a distinct dependence on filling factor and a unique response to temperature. We present a quantitative account of the observed structure, and argue it results from the repulsive Coulomb interactions between the tunneling electron and states localized at disorder potential wells. The quenching of the kinetic energy by the applied magnetic field leads to an electron addition spectrum that is primarily determined by the external magnetic field and is nearly independent of the disorder potential. Using a Hartree-Fock model we reproduce the salient features of the observed structure.**

The Quantum Hall Effect (QHE) has provided a rich variety of phenomena driven by the interacting nature of electrons. The application of a strong magnetic field perpendicular to a two dimensional electron gas (2DEG) quantizes the orbital degrees of freedom, which leads to the creation of highly degenerate Landau levels. Within each Landau level the kinetic energy is completely quenched, enhancing the importance of *e-e* interactions and giving rise to phenomena such as textured charge density [1] and the Fractional Quantum Hall Effect [2]. Disorder is a crucial ingredient in the physical description of the QHE, and is essential in accounting for the observation of quantized Hall conductance [2].

Recently, measurements of the tunneling density of states (TDOS) into the bulk of a QHE system were reported by Dial *et al.* using time domain capacitance spectroscopy [3]. The measurements revealed a new set of TDOS features ("sashes"), comprised of high energy bands, whose bottom decreases linearly with density. These observations led to some excitement, as one of the suggested explanations of these features involved new type of excitations in the quantum Hall regime. Since the existence of such yet unseen excitations on the physics of the quantum Hall effect may be far reaching, it is important to rule out an explanation based on more mundane physics. Here we present a possible mechanism that quantitatively reproduces these observations based on standard quantum Hall physics in the presence of *e-e* interactions.

Three energy scales determine the measured spectrum: the cyclotron energy $\hbar\omega_C$, the disorder potential fluctuations $\Delta V_{disorder}$ and the Coulomb interaction between electrons, characterized by $e^2/\varepsilon l_B$. Here $\omega_C=eB/m^*$, $e$ the electron charge, $B$ the applied magnetic field, $m^*$ the effective mass, $\varepsilon$ the sample dielectric constant and $l_B = \sqrt{\hbar/eB}$ the magnetic length (with $\hbar = h/2\pi$, $h$ being Planck constant). The picture described below is valid under strong magnetic fields, and assumes $e^2/\varepsilon l_B, \Delta V_{disorder} < \hbar \omega_C$. The electron density is described by the filling factor $\nu=BA/\phi_0$, where A is the sample area and $\phi_0=hc/e$, the quantum flux ($c$ is the velocity of light).

For simplicity, consider first the low density regime. When no electrons reside in the system ($\nu=0$), the TDOS describe the available single-particles states in the system, with each state doubly degenerate due to spin. Following the convention used in [3], where the Fermi level $E_F$ is set to $E=0$, it thus gives rise to an $E>0$ band, corresponding to the empty, disorder-broadened Landau level (*I* in Fig. 1a). This band will persist for higher filling factors, but the top of the band will decrease linearly, as $E_F$ rises with $\nu$. For small $\nu$, electrons are spatially separated, each sitting at a local minimum of the disorder potential. Thus two new features appear at the TDOS: removing an electron from an occupied state give rise to an $E<0$ band (*II* in Fig. 1a), and adding a second electron (of opposite spin) to an occupied state, with the cost of an additional Coulomb energy $\Delta E_{00}$, effectively creates a 'second Hubbard band' (*H1* in Fig. 1a). Since the kinetic energy is quenched by the strong magnetic field, the wavefunctions of these occupied states span an area of $\sim l_B^2$, thus $\Delta E_{00} \sim e^2/\varepsilon l_B$. Since $\Delta E_{00}$ is independent of density for high magnetic fields, the high energy band should thus replicate the band of occupied states, shifted by the constant energy $\Delta E_{00}$. The bottom of the higher Hubbard band should extrapolate to zero at $\nu=1$, since there all the single particle states are occupied and the Fermi energy lies at the bottom of the second Hubbard band, and the whole TDOS structure for $\nu<1$ is reflected about the E=0 axis for $\nu>1$ due to particle-hole symmetry. We identify the entire gap to the second Hubbard band as the '$\nu=1$ sash' reported in [3].

In principle, this description should be augmented by the physics of exchange and correlations which become relevant as $\nu$ increases. Close to $\nu=1$, the approximation of large separation between adjacent electrons is invalid, and a strong exchange gap [4-6] arises between the two spin types (*III* in Fig. 1a). The energy to add a neighboring same-spin electron becomes lower, leading to an abrupt upshift of the TDOS of the minority spins. We claim that the ingredients described above and modeled in detail below are enough to understand the experimental observations, indicating that electronic correlations, leading for example to the fractional quantum Hall effect, are not necessary to explain the experiment. An additional feature observed in the tunneling measurements is a gap near $E=0$. This gap arises from effective impedance to the tunneling, due to the required redistribution of charge in the system in order to allow for the injection of an electron [7-9]. This effect is well understood, and will not be modeled here.

The characteristic gap to the 2$^{nd}$ Hubbard band can be estimated as follows. Assuming the low-energy structure near a minimum of the potential is cylindrically symmetric, in the limit of strong magnetic field the associated electronic wavefunctions will be the $L_z$-eigenstates of the magnetic Hamiltonian, with $L_z$ the angular momentum projection in the direction of the applied field. The first electron entering the well will occupy the $L_z=0$ state, which has the lowest spatial span and hence the lowest energy. Once this state is occupied, injecting an opposite spin electron into the $L_z=0$ states involves a Coulomb energy cost arising from direct ('Hartree') interaction. An estimate of this energy can be obtained by calculating the interaction between two $L_z=0$ eigenstates, for a 2DEG well width of 23nm [3] and assuming the screening length is larger than the magnetic length. We find $\Delta E_{00} \approx 3.5 meV$ for B=4T, in excellent agreement with the measurements [3]. Furthermore, this energy gap scales as $B^{1/2}$, in agreement with the reported observation.

Additional lower energy bands may be created, arising from the injection of electrons into $L_z>0$ eigenstates in the well (H2 in Fig. 1b). Experimental observation of such features requires that the energy cost associated with this tunneling process be larger than the disorder broadening. Importantly, the short distance between the tunneling electrode and the 2DEG provides strong screening [3] and acts to lower the interaction energy with higher $L_z$ states, as these are spatially more extended. Calculating the Coulomb interaction between two electrons at the $L_z=0$ and $L_z=1$ states gives $\Delta E_{01} \approx 2meV$ at 4T and $\Delta E_{01} \approx 2.8meV$ at 8T, again in excellent agreement with the measured feature termed the '$^2$CF sash' in Ref. [3]. The exchange interaction between different $L_z$ eigenstates is very weak, so that in the limit of small filling factors no discernable separation is expected between the secondary bands arising from different spins. An important attribute of the $\Delta E_{01}$ gap is its symmetry around $\nu=1/2$ (IV in Fig. 1b), reflecting the electron-hole symmetry in the majority spin band. Again, this property is observed in the measurements [3] ('$\nu=1/2$ sash'). In addition to the low energy bands, higher energy features can emerge, in principle, which correspond to the addition spectrum of an electron to a location that is already doubly occupied (H3 in Fig. 1b). These features are not observed in the measurements, probably due to their stronger dependence on the local disorder potential profile, smearing them beyond the experimental resolution limit.

The actual calculation of the TDOS into a two-dimensional system in the presence of strong magnetic field, disorder and *e-e* interactions is numerically challenging. Here we present a restricted Hartree-Fock calculation of the TDOS into a quasi-one dimensional system, and demonstrate that it is sufficient to reproduce the physics described above, when the width of the wire *L* corresponds to the quantum confinement of the wavefunction in the two-dimensional system (*L~50nm)*. The calculation is limited to states in the lowest Landau level, and the electronic wavefunctions are assumed to be eigenstates of the disorder-free

Hamiltonian (in the Landau gauge). Disorder is introduced in the model by randomly determining a background potential profile. Calculations involve averaging over many (several thousands) such potential instances, so as to avoid any disorder-specific effects. The model assumes periodic boundary conditions in the transverse direction, quantizing the separation between neighboring states to $\Delta_x=\phi_0/BL$, with $\phi_0$ the flux quantum (Fig. 2a and b). Since the wavefunction are not calculated self consistently, this approximation effectively assumes that the disorder profile varies only in one direction. The role of higher $L_z$ eigenstates is played by farther-away neighbors, giving rise to lower energy bands similarly to the qualitative description (See Fig. 2b and c). While this model should be used with caution, since the quantitative predictions derived by the one dimensional constraint may be important, the qualitative predictions shown below are, as already mentioned, in agreement with the picture presented earlier. The agreement with the experimental results demonstrates that fractional-QHE physics, resulting from electronic correlations, are not necessary to understand these particular features. Due to the quenching of the kinetic energy, the Hamiltonian contains only two terms, the potential due to the disorder potential and the exponentially screened Coulomb interactions.

Figure 3 presents the results of such a calculation. In Fig. 3a, the screening length is assumed equal to the magnetic length, strongly attenuating the Coulomb interactions between states at different sites and reproducing the first Hubbard band alone (dashed blue line). For longer screening lengths (4 times $l_B$ in Fig. 3b), two additional features are observed, corresponding to the described bands H2 and H3. Additional features resulting from considering farther away occupied sites are attenuated by screening and disorder.

An important ingredient in the reported results [3] is a strong temperature dependence of the high-energy features that are shown to vanish for temperatures significantly smaller than the energy gap $\Delta E_{00}$. Our model reproduces this strong temperature dependence: besides the trivial broadening of the Fermi function in the metal gate, temperature also affects the occupation of states in the 2DEG. Since the high energy features (the higher Hubbard band) depends on these occupations, they will be sensitive to small temperature variations. In order to make these arguments quantitative within our numerical approach, we express the TDOS by the tunneling current,

(1) $I(V) \propto \int [f(\varepsilon - eV) - f(\varepsilon)] \nu_{2DEG}(\varepsilon;T) \nu_M(\varepsilon) d\varepsilon$,

where $\nu_{2DEG}(\varepsilon;T)$ and $\nu_M(\varepsilon)$ are the density of states in the probed 2DEG and in the metallic electrode, respectively, $f(\varepsilon)$ is Fermi function, and $V$ is the bias applied between the metallic electrode and the 2DEG. We assume, for simplicity, that the density of states in the metallic electrode is independent of energy, while that of the 2DEG, as mentioned above, may depend on temperature, due to the interactions. To recast, temperature should influence the observed signal in two ways: First, because of the Fermi function population, the occupations of the 2DEG and the metallic electrode are modified from a sharp profile in which all states below the Fermi energy are populated and no states above it are (Fig. 4a) to a smooth occupation distribution (Fig. 4b). Second, since the DOS of the 2$^{nd}$ Hubbard band directly mirrors the occupied state distribution, it will be similarly broadened. Note that this smearing of the high energy band is *independent* of the energy gap $\Delta E_{00}$. While the high energy bands remain distinguishable when considering the temperature smearing arising from each of these mechanisms alone, their combined effect leads to the complete vanishing of these features.

Including temperature into the calculation, we indeed find that the high energy features are strongly smeared (red line in Fig. 4c). Allowing, in addition, for the temperature smearing of the metallic electrode we find the high-energy structure becomes completely indiscernible (green line in Fig. 4c), reproducing the reported observations [3].

To conclude, we have demonstrated that the features observed in the experiment can be explained using standard quantum Hall physics in the presence of interactions. In the process of preparing this manuscript an additional possible explanation for [3] was given by Macdonald [10].

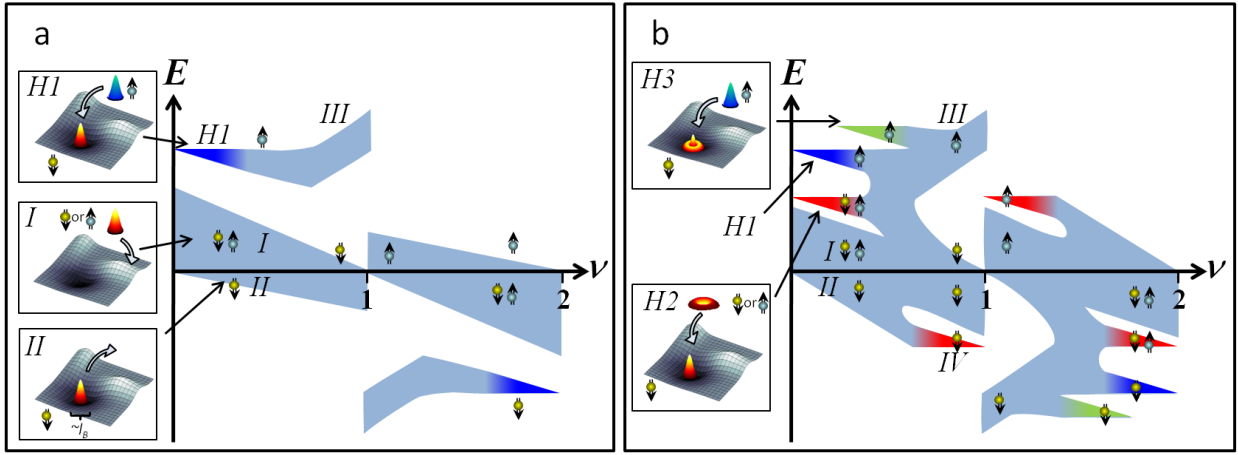

**Figure 1: (a)** Schematic of the TDOS dependence on the filling factor. For $\nu=1$, the TDOS is comprised of a disorder-broadened band (*I*). Energies are defined with respect to the chemical potential. For small filling factors, populated states (*II*) are comprised of electrons sitting at the ground state of disorder wells. Under strong magnetic field, the wavefunctions of the lowest energy states in a disorder potential well are only weakly dependent on details of the disorder potential. A 2$^{nd}$ Hubbard band is created (*H1*), resulting from the strong Coulomb interaction of adding an opposite-spin electron to an occupied state. Arrows indicate spin orientation. **(b)** Schematic of the TDOS dependence on the filling factor, including secondary bands arising from interactions with higher angular momentum eigenstates. In addition to the high energy band *H1*, a lower band (*H2*) arises from the injection of an electron into the next well eigenstate. A higher energy band (*H3*) is expected for larger filling factors, corresponding to the addition of an electron to a disorder potential well already occupied by two particles.

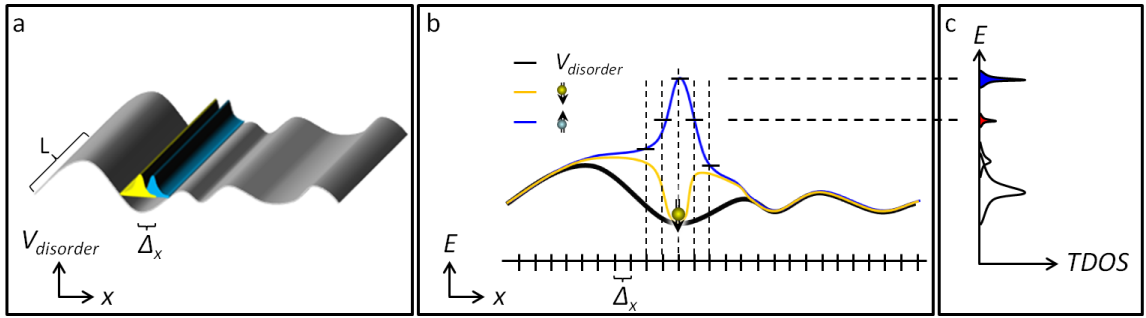

**Figure 2: (a)** We use a Hartree-Fock model to study the influence of Coulomb interactions on the tunneling density of states. **(b)** The background disorder potential (black line) gives rise to a broad tunneling density of states. When an electron occupies a potential well, the resulting Coulomb interaction creates high-energy bands for additional occupation the same site with an opposite spin electron, as well as occupation of nearby sites. **(c)** The resulting tunneling density of states is comprised of the low energy disorder-broadened band, and several high energy bands corresponding to the discrete energies to occupy neighboring states of an occupied site.

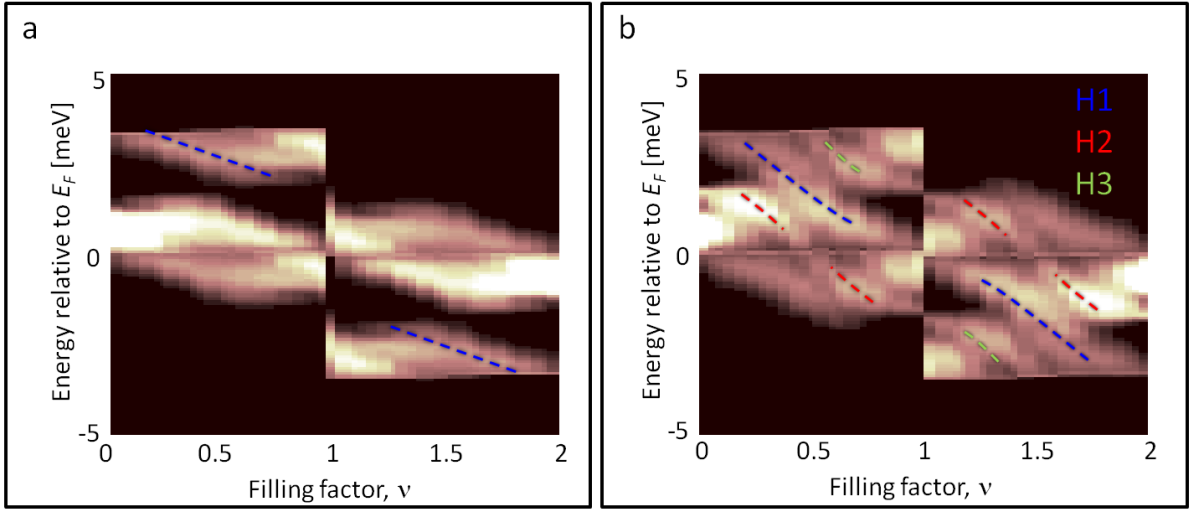

**Figure 3:** (a) A Hartree-Fock calculation of the TDOS, when the screening length is comparable to the magnetic length ($l_{sc} \approx l_B$). White features correspond to high TDOS. The strong feature marked by the dashed blue line is a 'second Hubbard band' created due to the added energy required to inject a particle into a disorder well already occupied by an opposite spin particle. (b) Calculated TDOS when screening length is larger than the magnetic length ($l_{sc} \approx 4 l_B$). Additional high energy features arise due to interactions between neighboring states (see text).

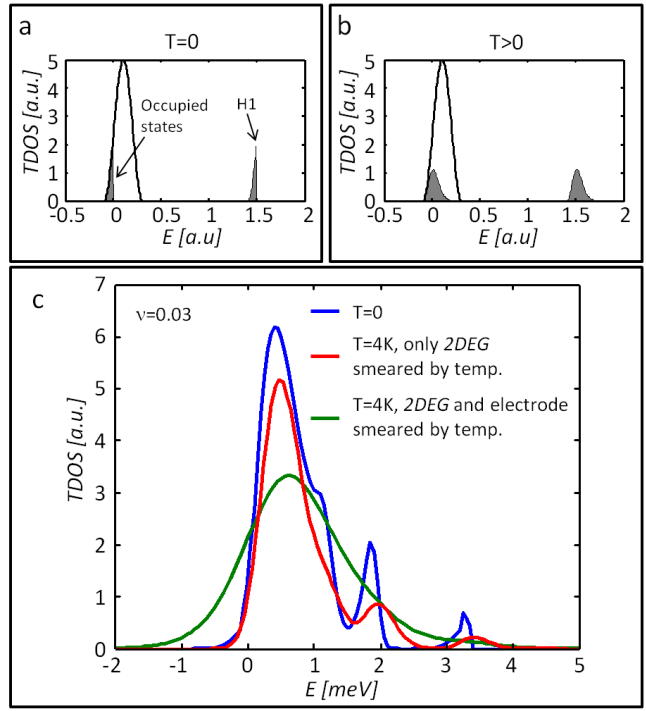

**Figure 4:** (a,b) Simplified model for the 2DEG density of states, presenting the non-trivial influence of temperature. (a) For T=0, all states below the Fermi level (set here to E=0) are occupied, giving rise to a duplicate high energy band *H1*. (b) For T>0, occupation of the low energy states is broadened. The high-energy $2^{nd}$ Hubbard band duplicates this broadening. (c) The calculated TDOS at T=0 (blue) for low filling factor of $\nu=0.03$. A temperature increase to 4K smears the TDOS structure, but the characteristic structure is still observable (red). Considering, in addition, the influence of the heated charge distribution in the tunneling electrode renders this structure undistinguishable (green).